# Ultra-thin silicon-on-insulator waveguide bend based on truncated Eaton lens implemented by varying the guiding layer thickness

S. Hadi Badri[a,*], M. M. Gilarlue[a], S. G. Gavgani[b]

[a] Department of Electrical Engineering, Sarab Branch, Islamic Azad University, Sarab, Iran

[b] Department of Electrical Engineering, Azarbaijan Shahid Madani University, Tabriz, Iran

* sh.badri@iaut.ac.ir

## Abstract

Silicon-on-insulator (SOI) waveguides with different geometries have been employed to design various integrated optical components. Reducing the bending radius of the SOI waveguides with low bending loss is essential in minimizing the footprint of light-wave circuits. The propagating mode is less confined in the core of the ultra-thin SOI waveguide and penetrates to substrate and cladding, leading to higher bending loss compared with conventional SOI waveguide with the thicker guiding layer. While various bending mechanisms have been utilized to reduce the bending loss of conventional SOI waveguides, the ultra-thin SOI waveguide bends have not been studied in detail. In this paper, we present a 60 nm-thick SOI waveguide bend based on the truncated Eaton lens implemented by varying thickness of the guiding layer. The three-dimensional full-wave simulations reveal that the designed waveguide bend, with a radius of 3.9 μm, reduces the bending loss from 3.3 to 0.42 dB at the wavelength of 1550 nm. Moreover, the bending loss for the wavelength range of 1260–1675 nm is lower than 0.67 dB while the bending loss in the C-band is lower than 0.45 dB.

## Keywords

Waveguide bend; Silicon-on-insulator waveguide; Ultra-thin SOI waveguide; Eaton lens

## 1. Introduction

Deviation of a waveguide from its straight path leads to diffraction of the light resulting in radiation losses and crosstalk to adjacent waveguides. Diverse methods have been introduced to reduce the bending radius while maintaining a low loss for conventional SOI waveguides. Mode converters [1] have been used to achieve a bending radius of 30 µm. A subwavelength grating waveguide bend with a radius of 10 µm has been reported [2]. A waveguide bend with optimized clothoid and normal curves has been studied [3]. A bending radius of 4 µm is achieved by this method. Modified Euler curves has been utilized to design waveguide bends with a radius of 45 µm [4]. A waveguide bend based on a Gradient index (GRIN) medium has been designed using transformation optics [5]. The designed waveguide bend has a large bending radius of 78 µm and has been implemented by varying the thickness of guiding layer. GRIN structures have also been used to design waveguide bends. The self-collimation property of a GRIN medium implemented by graded photonic crystal (GPC) has been used to bend light with a considerably large radius of $40a$, where $a$ is the lattice constant of the photonic crystal structure [6]. On the other hand, the refractive index at the interface of the straight waveguide and the bend ranges from 1.43 to 2.20. A GRIN medium designed with conformal mapping has been introduced as a waveguide bend and is implemented by GPC [7]. In this design, the refractive index at the interface of the straight waveguide and the bend ranges from 1.4 to 2.8. An antireflection coating has been used to reduce the reflection at the interface. For a waveguide with a width of 3.5 cm, a bending radius of 5.25 cm has been achieved.

SOI waveguides with different geometries have been utilized to implement various integrated silicon photonic devices. The silicon guiding layer in the ultra-thin SOI waveguides is thinner than 100 nm. The ultra-thin SOI waveguides have been employed to implement variety of devices such as modulators [8-10], delay lines [11], ring resonators [12], mode converters [13], sensors [14], spiral Bragg gratings [15], grating couplers [16, 17], multimode interference coupler [17], and Mach-Zehnder interferometers [17]. The propagating modes are confined in the guiding layer of a conventional SOI waveguide, consequently, these waveguides benefit from the high index-contrast between the core and claddings reducing the radiation loss in sharp bends. On the contrary, the fundamental mode's confinement in the guiding layer of the ultra-thin SOI waveguides is much lower than the conventional SOI waveguides. Furthermore, the propagating mode interacts weakly with the sidewalls [15] leading to higher radiation loss in the ultra-thin SOI waveguide bends. Therefore, designing a sharp ultra-thin waveguide bend is more challenging and is crucial in reducing the footprint of the components based on these waveguides.

Recently, novel applications have been introduced based on GRIN lenses such as Maxwell's fisheye [18, 19], Luneburg [20, 21], and Eaton [22, 23] lenses. The Eaton lens can bend light by 90°. Deflecting electromagnetic beam by the Eaton lens has been studied in the GHz range [24, 25]. Recently, we have designed a silica waveguide bend based on the Eaton lens implemented by GPC [22]. We have also studied the performance of a multimode polymeric waveguide bend using the Eaton lens implemented by multilayer structure [23]. In this paper, a compact bend is designed for an ultra-thin SOI waveguide based on the truncated Eaton lens. To the best of our knowledge, this is the first study that investigates the reduction of the bending radius of an ultra-thin SOI waveguide bend. The designed bend is implemented by varying thickness of the guiding layer of the slab waveguide. Our previously designed waveguide bends [22, 23] were evaluated with two-dimensional (2D) simulations while here we utilize three-dimensional (3D) simulations to validate our design. Numerical simulations indicate that the bending loss of the 60 nm-thick SOI waveguide can be substantially reduced from 3.3 to 0.42 dB. In order to emphasize the challenge of designing bends for the ultra-thin SOI waveguides, it should be noted that the bending loss of an SOI waveguide with a fixed 240 nm-thick guiding layer and the same bending curvature is merely 0.05 dB. The bending loss of the designed waveguide bend, with a bending radius of 3.9 μm, is lower than 0.45 dB in the C-band.

## 2. Bending by the Eaton lens

In this paper, we design a bend for an SOI waveguide with an ultra-thin guiding layer. The cross-section of the considered ultra-thin SOI waveguide is displayed in Fig. 1. An ultra-thin SOI waveguide bend has been studied in [17]. In order to compare our results with this study, we choose the same dimensions for the guiding layer. The width and thickness of the guiding layer are $w_{WG}$=950 nm and $t_{WG}$=60 nm, respectively. The silicon guiding layer is placed on the silica substrate and the upper cladding is air. The effective refractive index of this waveguide is calculated by Lumerical FDTD ® software [26] and is about 1.57 at a wavelength of 1550 nm. The contour plot of the electric field intensity of the transverse electric (TE) mode at 1550 nm is also shown in Fig. 1. In the TE mode, the electric field is perpendicular to the direction of propagation ($E_z$=0). The coordinate axes are also shown in Fig. 1.

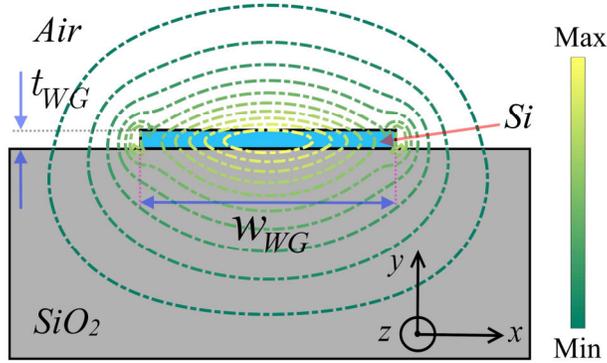

Fig. 1. The electric field intensity of TE mode at 1550 nm for an ultra-thin SOI waveguide with $w_{WG}$=*950* nm and $t_{WG}$=60 nm.

The Eaton lens can bend light by 90°, 180°, and 360°. In this paper, we employ the lens to bend light by 90°. The refractive index of the Eaton lens bending light trajectory by 90° is [22, 24]

$$n_{lens}^2 = \frac{R_{lens}}{n_{lens}r} + \sqrt{\left(\frac{R_{lens}}{n_{lens}r}\right)^2 - 1} \qquad (1)$$

where $R_{lens}$ is the radius of the lens, $r$ is the radial distance from the center, and $n_{lens}$ is the refractive index of the lens ranging from unity at its edge to infinity at its center. The conceptual design of the waveguide bend based on the Eaton lens is shown in Fig. 2. Ray-tracing calculations are performed with Comsol Multiphysics ® [27]. In this figure, the waveguide's core, with the effective refractive index of 1.57, is also shown. In order to minimize the reflection from the interface of the waveguide and the lens, the numerically calculated $n_{lens}$ is multiplied by 1.57. Since the refractive index of the lens approaches infinity at its center, we limited the maximum refractive index to 3.5 in Fig. 2(a) to make the refractive index distribution more perceptible. In this figure, the design parameters are $w_{WG}$=950 nm, $R_{lens}$=2250 nm, and $\delta$=1157 nm. $\delta$ is the distance between the center of the lens and the inner boundary of the waveguide. The ray-tracing calculations indicate that light is limited to a region shown by bolded lines in Fig. 2(a), therefore, we reduce the footprint of the bend by truncating the lens as shown in Fig. 2(b). Maximum refractive index of the truncated Eaton lens, $n_{max}$, depends on the distance between the center of the lens and the inner boundary of the waveguide, $\delta$. As illustrated in Fig. 3, for the lens with the fixed radius, $R_{lens}=2250$ nm, as $\delta$ decreases the maximum refractive index of the truncated lens increases. It should be noted that refractive index of the Eaton lens sharply increases as the radial distance from the center of the lens approaches zero. Therefore, the gradient of the refractive index of the truncated Eaton lens increases near the center of the truncated lens making it difficult to implement. We choose $\delta=R_{lens}-1.15\times H_{core}=1157$ nm to avoid sharper gradients of the refractive index while limiting the $n_{max}$ to a smaller value, i.e., 2.96. This eases the implementation of the lens which is discussed in the next section. The bending radius of the waveguide bend, $R_{bend}$, depends on the radius of the lens, $R_{lens}$. When $\delta$ is fixed, we can decrease $R_{bend}$ by reducing $R_{lens}$. However, changing $\delta$ while $R_{lens}$ is fixed results in small changes of $R_{bend}$ (smaller than 0.3 µm). On the other hand, the minimum value of $R_{lens}$ depends on the width of the waveguide ($w_{WG}$). The radius of the lens should be larger than $2\times w_{WG}$. In overall, $R_{lens}$ and $\delta$ are chosen considering the width of the waveguide while keeping $n_{max}$ to a feasible value.

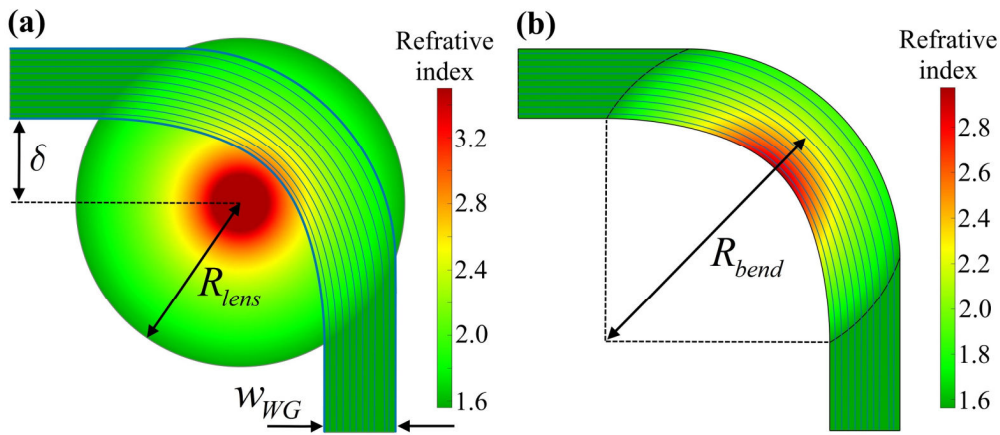

Fig. 2. Ray-tracing calculations for the conceptual design of the waveguide bend based on the Eaton lens. a) The refractive indices of the lens and the bent waveguide is displayed. B) The lens is truncated to reduce the footprint of the bend.

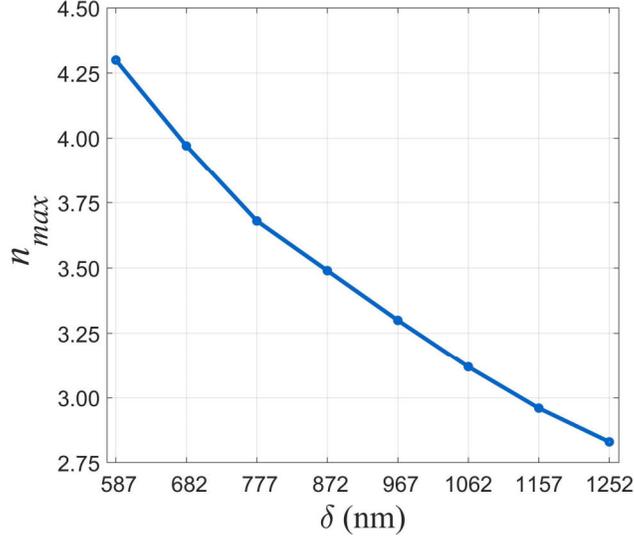

Fig. 3. Maximum refractive index of the truncated Eaton lens ($n_{max}$) vs the distance between the center of the lens and the inner boundary of the waveguide ($\delta$).

## 3. Implementing the lens by varying thickness of the guiding layer

GRIN media can be implemented by graded photonic crystals [6, 7, 28-33], multilayer structures [34-36], and varying thickness of the guiding layer in a slab waveguide [37-39]. In this section, we briefly explain the method to implement a GRIN medium by varying the thickness of the guiding layer in a slab waveguide. Various GRIN devices have been designed by varying the thickness of silicon layer and they have been fabricated using grayscale e-beam lithography [5, 38, 40]. The refractive index of the Eaton lens is shown in Fig. 4(a). The minimum and maximum refractive indices of the truncated Eaton lens are shown in this figure. In order to map this refractive index profile to the thickness of the silicon guiding layer, we should calculate the effective refractive index of the silicon slab waveguide. The effective refractive index of the slab waveguide depends on the absolute refractive index of the substrate, guiding layer, and upper cladding materials as well as the dimensions of the guiding layer. We use silica, silicon, and air as substrate, guiding layer, and upper cladding materials in the slab waveguide, therefore, by fixing the width of the guiding layer, the effective refractive index of the waveguide depends only on its thickness. Lumerical software is used to calculate the effective refractive index of the waveguide with a thickness ranging from 40 to 510 nm. Fig. 4(b) illustrates the fitted curve of the effective refractive index of the slab waveguide versus its guiding layer's thickness. To map the refractive index of the designed bend [Fig. 2(b)] to the thickness of the silicon layer, we follow these steps: we calculate the refractive index of the bend at each point from Fig. 4(a). For instance, the refractive index of the lens at a distance of $r=1.72$ μm from the center is $n=1.9$. Then we map this refractive index to the thickness of a slab waveguide based on the fitted curve of Fig. 4(b). In this figure, the minimum and maximum thicknesses of the silicon layer are also determined. The refractive index of $n=1.9$ corresponds to a slab waveguide with a thickness of $t=90$ nm as shown in Fig. 4(c). Consequently, the thickness of the silicon layer in the waveguide bend at the distance of $r=1.72$ μm from the center is $t=90$ nm which is shown in Fig. 4(d). This procedure is repeated for each point and the final structure of the waveguide bend is created with interpolation between these points. In Fig. 4(d), only the varying thickness of the waveguide bend is shown and the silica substrate and upper air cladding are not shown. To provide another point of view for the readers, the resulting structure is also shown in Fig. 5. As it is obvious in Fig. 4(b), a considerably large variation of the guiding layer thickness is required to implement refractive indices of larger than about 3. Achieving sharp gradients in the guiding layer thickness may be limited by the fabrication processes. Therefore, we chose the value of $\delta$ in a manner that the refractive index of the truncated Eaton lens is limited to values smaller than 3 and also avoid sharper gradients of the guiding layer thickness.

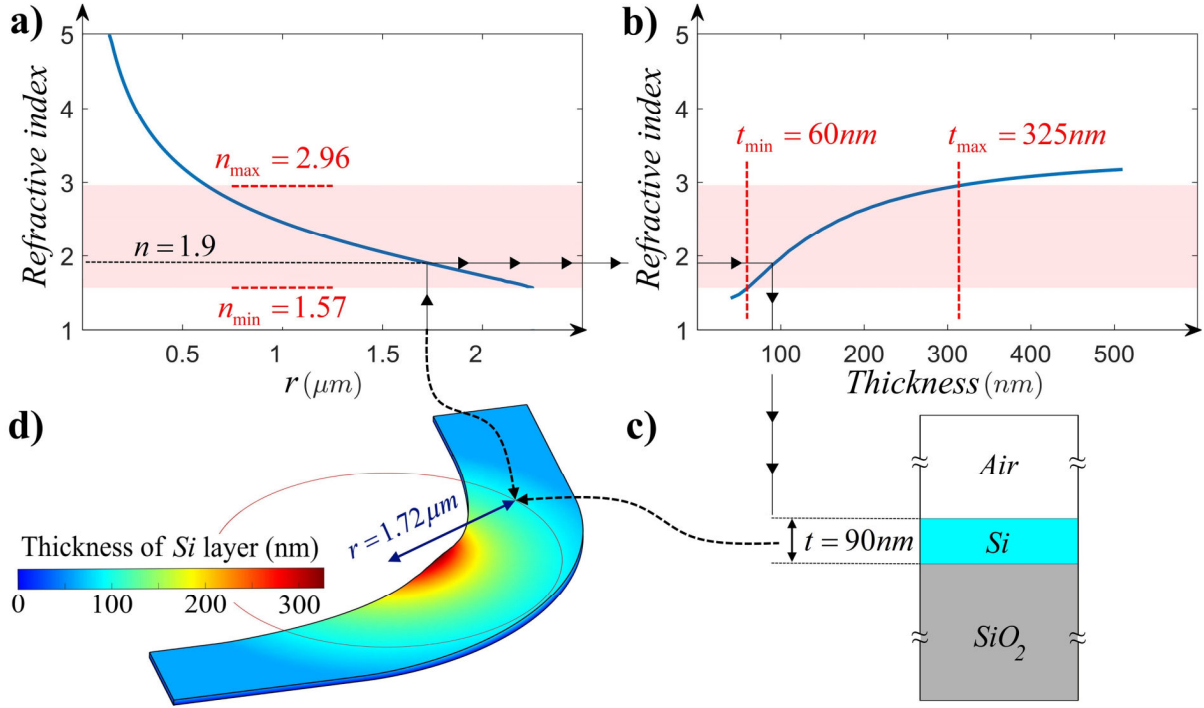

Fig. 4. Mapping the refractive index of the truncated lens to the thickness of the silicon guiding layer. a) Refractive index of the Eaton lens with respect to the radial distance from the center of the lens. b) Effective refractive index of the silicon slab waveguide with respect to the guiding layer's thickness. c) Silicon slab waveguide corresponding to effective refractive index of 1.9. d) The refractive index of the designed bend of Fig. 2(b) mapped to the thickness of the silicon layer.

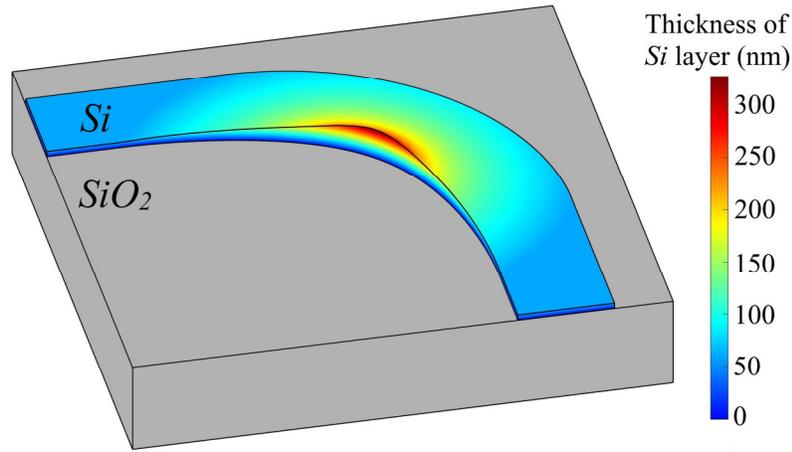

Fig. 5. The designed refractive index of the bend [Fig. 2(b)] is translated to the silicon guiding layer's thickness. The upper air cladding is not shown.

## 4. Results and discussions

We employed 3D finite-difference time-domain (FDTD) to evaluate the performance of the designed waveguide bend. The built-in material models of the Lumerical software are used in simulations. The "custom non-uniform" method is used for meshing in the Lumerical. The maximum meshing steps are chosen based on the size and geometry of the structure. Since the width of the waveguide is 950 nm and considering the gradual thickness change in the bent waveguide, we choose the maximum mesh size in the lateral direction as 30 nm. However, the thickness of the waveguide is 60 nm so in the vertical direction the maximum mesh size is chosen as 5 nm. The velocity vectors and contour plots of the electric field intensity of the TE mode propagating through a simple bent waveguide is displayed in Fig. 6(a). The bending loss of the bend is 3.3dB at the wavelength of 1550 nm. The electric field leaks out of the waveguide bend without the lens. A waveguide bend based on the complete Eaton lens is displayed in Fig. 6(b). In this case, the bending loss decreases to 1.7 dB. The complete Eaton lens bends the electromagnetic wave, however, it is partially successful in guiding the wave towards the output waveguide. By truncating the Eaton lens, the bend is surrounded by a lower refractive index of air and consequently the electromagnetic field is confined inside the bend resulting in considerable reduction of bending loss. Moreover, the bent waveguide designed by the truncated Eaton lens occupies a smaller space. The bending loss of the waveguide bend based on the truncated Eaton lens is 0.42 dB at the wavelength of 1550 nm. As shown in Fig. 7, the out-of-plane confinement increases in the bent waveguide due to the thickening of the guiding layer which in turn increases the effective refractive index of the waveguide. This figure is created with Comsol.

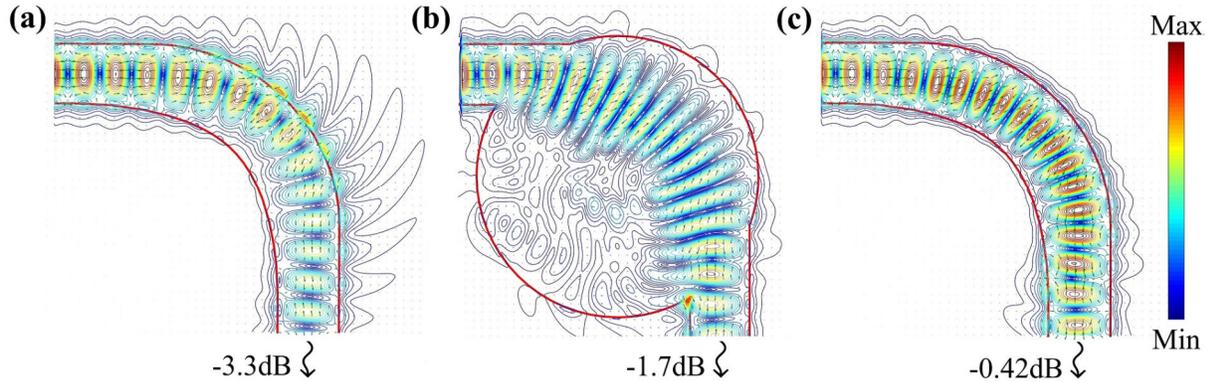

Fig. 6. The vectors and contour plots of the electric field intensity of the TE mode light at the wavelength of 1550 nm propagating through the waveguide bend a) without the lens, b) with complete Eaton lens, and c) the truncated Eaton lens.

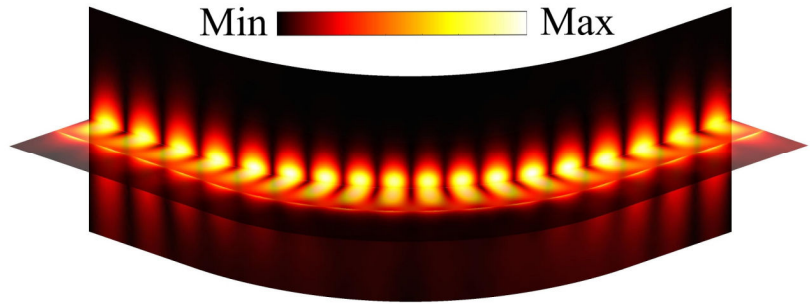

Fig. 7. Propagation of the TE mode light at 1550 nm through the designed bend.

The bending loss as a function of wavelength is shown in Fig. 8. The bending loss of the designed waveguide based on the truncated Eaton lens is lower than 0.67 dB in the entire optical communication bands. Moreover, the bending loss is lower than 0.45 dB in the C-band. The bending loss of the waveguide without the help of the lens is noticeably high in longer wavelengths. In this case, the bending loss is lower than 5.26 dB and higher than 0.73 dB in the 1260-1675 nm bandwidth. We also simulated a waveguide bend with the same bending curvature while fixing the thickness of the silicon guiding layer to 240 nm. In this case, the bending loss is 0.05 dB at 1550 nm. This is due to the fact that the electromagnetic field is largely confined in the thick guiding layer, therefore, high contrast between the core and cladding helps to confine and bend the field efficiently. On the contrary, in the 60 nm-thick waveguide, the electromagnetic field penetrates deeper into the substrate and cladding leading to higher bending loss similar to low-index contrast waveguides. As the wavelength of the input light increases a larger portion of the field extends to the substrate and cladding, consequently, the bending loss increases as is evident in Fig. 8. The fabrication imperfections, such as deviation in the thickness of the designed lens and sidewall roughness of the bent waveguide, are inevitable. Hence, we numerically estimate the effect of these imperfections on the performance of the bent waveguide. To this end, the spatial Si thickness randomly ranging from −20 to +20 nm is added to the designed thickness of the ideal lens. Furthermore, a random sidewall roughness of ±20 nm is also added to the model. These imperfections may increase the bending loss by up to 0.3 dB in the C-band.

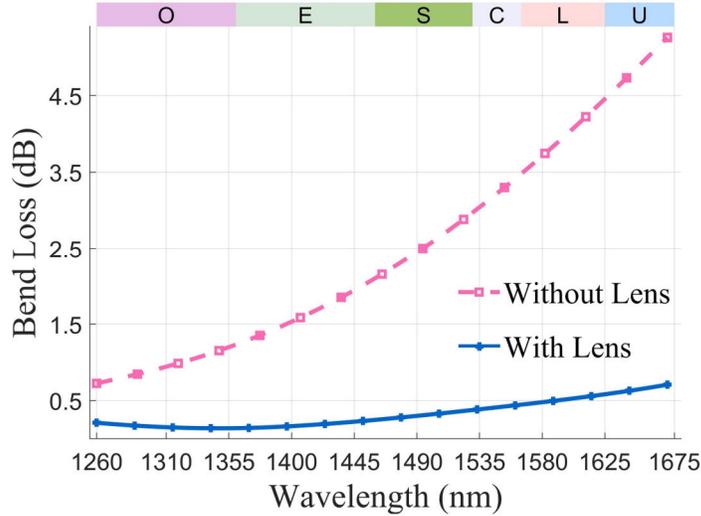

Fig. 8. The bending loss for the waveguide bend with and without the help of the truncated Eaton lens.

Finally, we compare our results with previously designed waveguide bends in Table 1. The waveguide type, bending mechanism, bending radius, bending loss, bandwidth, and evaluation method are compared in this table. When available both numerical and experimental results are presented. To distinguish between the numerical and experimental results in the table, we present the numerical results in the parentheses. Compared to the previous studies, our design has a small bending radius while its bending loss is in an acceptable range. As mentioned before, the propagating mode is partially confined in the guiding layer of the ultra-thin SOI waveguide leading to higher bending loss compared to the conventional SOI waveguides. It should be noted that our results are based on numerical simulations while other studies report measurement results. In references [6, 7], an electromagnetic field in the free space is bent with the GRIN media implemented by GPC. References [6, 7] report transmission versus the normalized frequency based on the lattice constant. Since the lattice constant is not given in these references, we choose the lattice constant corresponding to the center of the photonic bandgap or the passband. In reference [7], the width of the waveguide is 3.5 cm and its bending radius is 5.25 cm. We should also point out that our proposed design is more suitable for ultra-thin SOI waveguides. In order to apply our design to thicker waveguides, the radius of the Eaton lens as well as the distance from the center of the lens, δ, should be increased leading to considerable increase in the effective bending radius.

Table 1. Comparing the designed bend with previous studies.

| [Ref.] year | Waveguide Type | Bending mechanism | Bending radius (μm) | Bending loss (dB) @1550 nm | Bandwidth (nm) | Evaluation method |
|---|---|---|---|---|---|---|
| [1] 2018 | SOI | Mode-converter | 30 | 0.66 (0.01) | 1520-1600 (1500-1600) | Experimental (Numerical) |
| [2] 2019 | SOI | Subwavelength gratings | 10 | 0.2 (<0.2) | 1520-1600 (1500-1600) | Experimental (Numerical) |
| [3] 2017 | SOI | Clothoid and normal curves | 4 | 0.002 | 1520-1620 | Experimental |
| [4] 2018 | SOI | Modified Euler curves | 45 | 0.25 (<0.1) | 1520-1610 (1500-1600) | Experimental (Numerical) |
| [5] 2012 | SOI | GRIN medium | 78.8 | 2.6 (2.5) | - | Experimental (Numerical) |
| [6] | Free space | GRIN medium | 29 | (0.04) | (1375-1770) | (Numerical) |
| 2013 | | | | | | |
| [7] 2011 | Free space | GRIN medium | 5.25 cm | (0.08) | (1150-3250) | (Numerical) |
| [17] 2015 | Ultra-thin SOI | - | 30 | 0.01 | 1520-1600 | Experimental |
| This work | Ultra-thin SOI | Eaton lens | 3.9 | (0.42) | (1260-1675) | (Numerical) |

**Conclusion**

Ultra-thin SOI waveguides are employed in various components, however, the propagating mode is partially confined in the thin guiding layer leading to higher bending loss compared to the SOI waveguide with the thicker guiding layer. Numerical simulations reveal that the bending loss can be decreased from 3.3 to 0.42 dB by utilizing the truncated Eaton lens for the 60 nm-thick SOI waveguide. We implement the lens by varying thickness of the guiding layer. The bending loss of designed ultra-thin SOI waveguide bend, with an effective bending radius of 3.9 μm, is lower than 0.67 dB in the entire O, E, S, C, L, and U bands of optical communication.